\documentclass[twocolumn]{aastex631}

\newcommand\emin{\epsilon_{\rm min}}

\newcommand\gammae{\gamma_{e,\rm max}}

\newcommand\me{m_{e}}
\newcommand\sT{\sigma_{\rm T}}
\newcommand\rg{r_{\rm g}}

\newcommand\sgg{\sigma_{\gamma\gamma}}

\newcommand\Bh{B_{\rm H}}

\newcommand\Lcur{L_{\rm cur,pk}}
\newcommand\Lic{L_{\rm IC,pk}}

\newcommand\Lbz{L_{\rm BZ}}

\newcommand\nism{n_{\rm ISM}}
\newcommand\m{m_{p}}

\newcommand\hism{h_{\rm ISM}}
\newcommand\vavg{v_{\rm avg}}
\newcommand\Rh{R_{\rm h}}
\newcommand\Mh{M_{\rm h}}

\newcommand\csism{c_{s,\rm ISM}}
\newcommand\Medd{\dot{M}_{\rm Edd}}
\newcommand\cum{N(>\!\! F)}

\newcommand\tggmad{\tau_{\gamma\gamma, \rm MAD}}
\newcommand\Ndet{N_{\rm det}}

\usepackage{amsmath,amssymb}

\shorttitle{Spark gap IBHs as possible GeV-TeV un-IDs}
\shortauthors{Kin et al.}
\begin{document}

\title{Galactic Isolated Stellar-Mass Black Holes with the Magnetospheric Spark Gap\\ as Possible GeV-TeV Gamma-ray Unidentified Sources}

\author[0000-0002-9712-3589]{Koki Kin}
\affiliation{Astronomical Institute, Graduate School of Science, Tohoku University, Sendai 980-8578, Japan}

\author[0000-0002-5916-788X]{Riku Kuze}
\affiliation{Astronomical Institute, Graduate School of Science, Tohoku University, Sendai 980-8578, Japan}

\author[0000-0003-2579-7266]{Shigeo S. Kimura}
\affiliation{Astronomical Institute, Graduate School of Science, Tohoku University, Sendai 980-8578, Japan}
\affiliation{Frontier Research Institute for Interdisciplinary Sciences, Tohoku University, Sendai 980-8578, Japan}





\begin{abstract}
Billions of isolated stellar-mass black holes (IBHs) are thought to wander through the interstellar medium (ISM) in the Galaxy, yet only one has been detected.
IBHs embedded in ISM would accrete gas via Bondi-Hoyle-Littleton accretion, and with efficient magnetic flux accumulation, the magnetosphere would be formed in the vicinity of IBHs.
We explore the detectability of such IBHs through high-energy gamma rays from spark gaps in their magnetospheres based on our recent numerical simulation.
The gap gamma rays can be bright at the GeV-TeV energies when IBHs are in the dense ISM.
About $10^3$ and $10$ IBHs might be contained in unidentified objects of the \textit{Fermi} Large Area Telescope and the High Energy Stereoscopic System, respectively.
A future Galactic plane survey by the Cherenkov Telescope Array Observatory would lead to $\sim10^2$ detections.
We also evaluate the combined gamma-ray emission of IBHs in the Galaxy and find that the IBHs may contribute to the Galactic diffuse gamma rays.
IBHs will emit optical and X-ray photons from their accretion disk as counterparts, potentially useful for identifying candidates.

\end{abstract}



\section{Introduction} \label{sec:intro}
The existence of stellar-mass black holes (BHs) has been confirmed through decades of observations of Galactic X-ray binaries \citep[see][]{BlackCAT, WatchDog}, increasing numbers of gravitational wave detections from compact binary mergers \citep{Abbott23}, and the analysis of the dynamical motion of binaries in \emph{Gaia} data \citep[Gaia BH 1, 2, and 3;][]{ElBadry23a, ElBadry23b, GaiaBH3}.
$10^8$-$10^9$ stellar-mass BHs are thought to exist in the Galaxy as the end products of massive ($\gtrsim25M_\odot$) stars \citep[e.g.][]{Shapiro86, Caputo17}, and the governing fraction of them are expected to be isolated stellar-mass BHs (IBHs) \citep[][]{Olejak20}.
However, we have not detected any hints of electromagnetic signatures related to IBHs, except for one microlensing event \citep[OGLE-2011-BLG-0462/MOA-2011-BLG-191;][]{Sahu22, Lam22}.

IBHs around the Galactic plane would accrete interstellar medium (ISM) via the Bondi-Hoyle-Littleton (BHL) accretion \citep[see][for a review]{Edger04}.
The non-zero net angular momentum of infalling gas will force them to form the accretion disks around IBHs (\citealp{Shapiro76, Fujita98, Agol&Kamionkowski02, Ioka17}).
In a typical ISM environment, the BHL accretion rate onto IBHs is so low that their accretion disk would be a radiatively inefficient accretion flow \citep[RIAF;][]{Ichimaru77, Narayan94, Yuan14}.
The accreting gas in RIAF is hot, which leads them to achieve a high radial velocity.
As a result, an efficient accumulation of the magnetic flux onto IBHs is expected \citep{Cao11, Barkov12, Ioka17}.
The disk near the IBH will thus be the magnetically arrested disk \citep[MAD;][]{Narayan03, Narayan12}, in which the magnetic flux threading the central BH is at a saturated level.
Recent general relativistic magnetohydrodynamic (GRMHD) simulations \citep{Kaaz23, Kwan23, Galishnikova24, Kim24} support this scenario.
The detectability of the IBHs via multi-wavelength emission from accretion disks, disk-induced outflows, and jets has been investigated in many literatures \citep[e.g.,][]{Barkov12, Fender13, Tsuna18, Tsuna19, Kimura21}.

At the inner edge of the MAD, accretion is truncated by the magnetic pressure, and the magnetosphere is formed.
We have recently investigated the plasma dynamics in the magnetosphere of the stellar-mass BHs using one-dimensional general relativistic particle-in-cell (GRPIC) simulations \citep{Kin24}.
We have shown that the longitudinal electric field (i.e. the spark gap) emerges during the charge-starved state of the magnetosphere, as is the case for several previous research \citep[see, e.g.,][]{Levinson18, Kisaka20, Crin20, Chen20, Kisaka22, Hirotani22}.
The spark gap efficiently accelerates the electrons, emitting high-energy gamma rays.
We have also studied the range of the mass accretion rate that brightens gamma-ray emission.
Our analysis has suggested that gamma rays escaping from IBH magnetospheres embedded in the dense ISM are bright enough for detection by the \emph{Fermi} Large Area Telescope \citep[LAT;][]{F-LAT}, the High Energy Stereoscopic System \citep[H.E.S.S.;][]{CrabwHESS}, and the Cherenkov Telescope Array Observatory \citep[CTAO;][]{CTA}).

In this work, we further explore this scenario and thoroughly discuss the detection prospects of signals from MADs/spark gaps of IBHs.
The paper is organized as follows: Section~\ref{sec:mth} gives a method for our calculations.
Then, we show the results of the expected number, physical properties, and spectral properties of detectable IBHs in Section~\ref{sec:res}.
Section~\ref{sec:disc} discusses the identification strategy and caveats.
Finally, conclusions are given in Section~\ref{sec:conc}.
We have also performed the 1D GRPIC simulations with various BH spin values to investigate the dependence of gamma-ray luminosity on the spin.
Simulation results are summarized in Appendix \ref{appB}.
Here and hereafter we use the convention $Q_x=\left(Q/10^x\right)$ in cgs units, except for the BH mass, $M = 10^x M_x M_\odot$.

\begin{deluxetable*}{cccccc}
\tablenum{1}
\tablecaption{Physical parameters of ISM. $\nism$ is the reference value of the ISM number density, $c_{\rm s, ISM}$ is the effective sound velocity, $\hism$ is the pressure scale height, $\xi(R_{\rm obs})$ is the calculated volume filling factor at the Sun's location, and $\dot{m}(M_1,v_6)$ is the Eddington ratio calculated using $M=10M_\odot$, $v= 10\,\rm km\,s^{-1}$, and reference values of $\nism$ and $\csism$.}
\label{tab:ISMparam}
\tablewidth{0pt}
\tablehead{
ISM Phase & $\nism\,[\rm cm^{-3}]$ & $c_{\rm s, ISM}\,[\rm km\,s^{-1}]$ & $\hism\,[\rm kpc]$ & $\xi(R_{\rm obs})$ & $\dot{m}(M_1,v_6)$ }  
\startdata
 Molecular Clouds & $10^2, 10^3$ & $3.7(\nism/100\,\rm{cm^{-3}})^{-0.35}$ & $0.075$ & $3.9\times10^{-3}$, $6.2\times10^{-5}$ & $3.2\times10^{-2},\,2.8\times10^{-3}$\\
 Cold HI & $10$ & 10 & 0.15 & 0.061 & $1.2\times10^{-4}$\\
 Warm HI & 0.3 & 10 & 0.5 & 0.37 & $3.6\times10^{-6}$\\
\enddata
\end{deluxetable*}
\section{Method}\label{sec:mth}
\subsection{Model Overview}
\begin{figure}[t]
\hspace*{0.5cm}
 \includegraphics[keepaspectratio, trim = 0 30 0 0, scale=0.6]{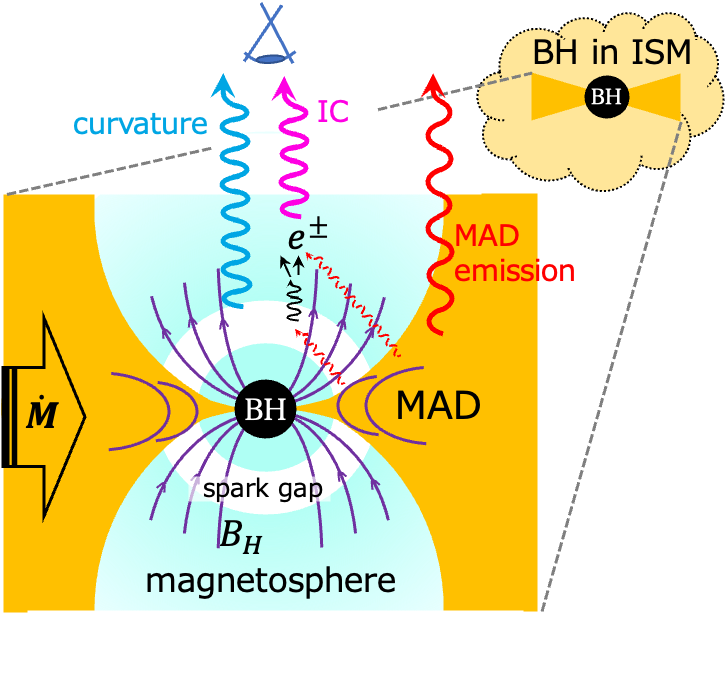}
\caption{Schematic image of the IBH-MAD-magnetosphere system.}
\label{sche}
\end{figure}
We consider Galactic IBHs wandering through the ISM.
Fig.~\ref{sche} provides the schematic image of our model.
IBHs accrete ISM and the accretion disk is formed.
The disk near IBHs will be the MAD due to efficient magnetic flux accumulation, and then the magnetosphere is formed around the IBHs.

The BHL mass accretion rate for each IBH is estimated as a function of the BH mass, $M$, the BH proper motion velocity, $v$, and the ISM number density, $\nism$, 
\begin{equation}
 \begin{aligned}
 \label{acc_rate}
 &\dot{M}(M,\,v,\,\nism)=\lambda_{w}\dfrac{4\pi G^2M^2\mu\m\nism}{V_{\rm rela}^3}\\
 &\hspace{20pt}\simeq1.7\times10^{14}\lambda_{w,0}M_1^2\left(\dfrac{\mu}{1.26}\right)n_{\rm ISM,1}V_{\rm rela,6.3}^{-3}\mathrm{g\,s^{-1}},
 \end{aligned}
 \end{equation}
where $\m$ is the proton mass, $\mu$ is the mean molecular weight, $G$ is the gravitational constant, $\lambda_{w}$ is the wind mass loss rate, and $V_{\rm rela}=(\csism^2+v^2)^{1/2}$ is the IBH relative velocity considering the effective sound velocity of ISM, $\csism$.
We focus on $\dot{M}$ in the sub-Eddington regime, in which we expect a MAD formation: $\dot{m}\equiv\dot{M}/\Medd\leq10^{-2}$, where $\Medd=4\pi GM\m/\sT c\simeq1.4\times10^{18}M_1\,\mathrm{g\,s^{-1}}$ is the Eddington mass accretion rate. ($\sT$ is the Thomson cross section and $c$ is the speed of light.)
This condition is satisfied for most of the parameter ranges in this work. (compare Eq.~(\ref{acc_rate}) with $\Medd$.)
We then consider emissions from two regions: the MADs and the spark gaps in the magnetospheres.
MAD multi-wavelength emission properties are sensitive to $\dot{M}$.
Gap gamma-ray emission properties are also determined by $\dot{M}$, and additionally, by the BH spin, $a$ (see Section \ref{ss:gapf}).

In the actual calculation, we first determine the spatial and the velocity distribution of IBHs in the Galaxy through the simplified dynamical calculation (Section~\ref{ss:dyn_cal}).
Each IBH's spatial correlation probability between ISM i.e. the volume filling factor, $\xi$, can be derived as a function of $\nism$ and IBH's position in the Galaxy, considering the ISM's spatial distribution.
Then, we give each IBH a mass and a spin (Section~\ref{ss:mds}).
The IBH-MAD/spark gap multi-wavelength emission spectra for a given $\nism$ are calculated by using those parameters (Sections~\ref{ss:flux}, \ref{ss:gapdyn}, and \ref{ss:gapf}).
The observed flux of each IBH is calculated based on its position in the Galaxy.
We discuss the possible reduction of the persistent gamma-ray flux due to the expected time-dependency of the spark gap in Section~\ref{ss:reduc}.

\subsection{IBH Distribution Calculation}\label{ss:dyn_cal}
We calculate the distribution of BHs in 2D cylindrical coordinates perpendicular to the Galactic plane, $(R,\,z)$, following the method described in \citet{Tsuna18}.
To reduce the computational time, we calculate $10^5$ IBHs and scale the result to the estimated IBH total number in the Galaxy, $N_{\rm tot}=10^8$ (see Section~\ref{res1}).
One important parameter for the calculation is the IBH initial velocity, mostly determined by the kick from the supernova explosion.
We assume the kick velocity distribution as the 3D Maxwell-Boltzmann distribution with a mean kick velocity $\vavg$.
We run the calculation for $\vavg=\{10, 50, 100, 400\}\,\rm km\,s^{-1}$, which are consistent with the range of the kick velocity estimated from the BH X-ray binary distribution in the Galaxy \citep[e.g.][]{Repetto17, Nagarajan24}.
We note that we have introduced the smoothing terms in the equations of motion to reduce numerical errors on the calculated BH velocity.
More details for the calculation and the resultant spatial/velocity distributions for different values of $\vavg$ are shown in Appendix \ref{appA}.
Overall results are consistent with \citet{Tsuna18}: the IBH spatial distribution approaches from the initial exponential disk shape to a more spherical shape for $\vavg\geq100\,\rm km\,s^{-1}$.
In contrast, the exponential disk profile is preserved for lower $\vavg$ cases.

\subsection{ISM Property, Distance, Mass, \& Spin}\label{ss:mds}
We consider three ISM phases of different temperatures and the density \citep[][]{ISMref}, which are molecular clouds, cold neutral medium (cold HI), and warm neutral medium (warm HI).\footnote{We have checked that IBHs in warm H$\rm I\hspace{-.1em}I$ and hot H$\rm I\hspace{-.1em}I$ could not achieve high $\dot{M}$ enough to contribute to the detection.}
Parameters are enlisted in Table~\ref{tab:ISMparam}.
The ISM volume filling factor for a number density of the certain ISM phase at a given position is described as $\xi(\nism;\,R,\,z)=\int d\xi/d\nism(\nism;\,R)H(\hism-z)d\nism$, where $H$ denotes the Heaviside function.
The differential fraction $d\xi/d\nism$ is obtained by considering the $\nism$ dependence of $\xi$ for molecular and cold HI gas, and the observed cylindrical distribution of the ISM column density in the Galactic disk \citep[see][for details]{Agol&Kamionkowski02, Tsuna18}.

\begin{figure*}[t]
\includegraphics[keepaspectratio, trim = 5 5 0 0, scale=0.42]{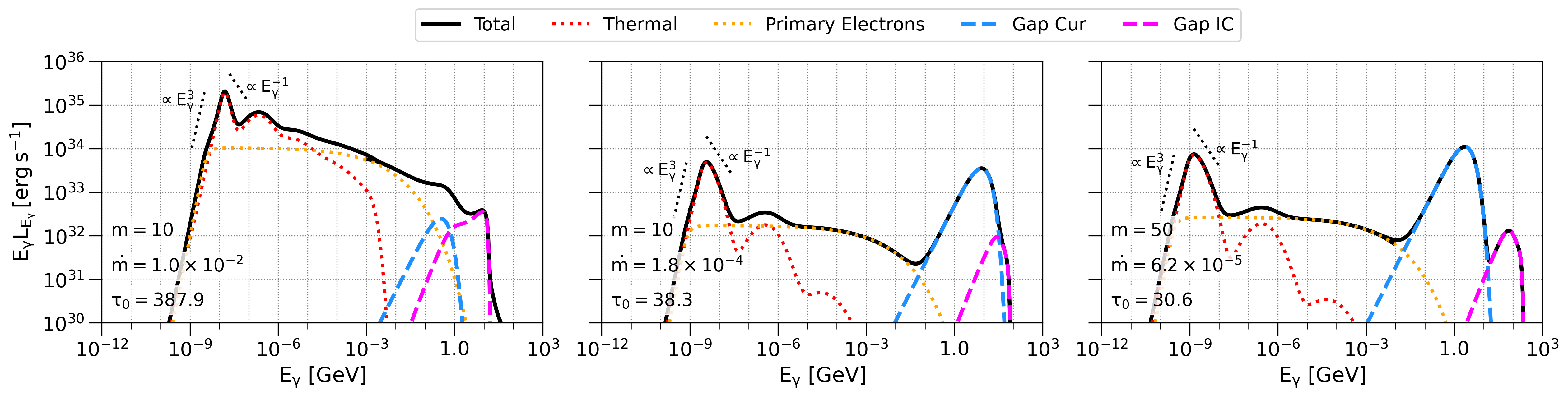}
\caption{The multi-wavelength luminosity spectra of the IBH-MAD/spark gap for $a=0.9$. Each panel shows the results for $(m=M/M_\odot,\,\dot{m})=(10,\,1.0\times10^{-2}),\,(10,\,1.8\times10^{-4}),\,(50,\,6.2\times10^{-5})$. The black solid line denotes the total spectrum. The dotted lines show synchrotron emission from MAD thermal electrons (red) and non-thermal primary electrons (orange). The dashed lines show spark gap curvature radiation (cyan) and IC up-scattering emission components (magenta).}
\label{spectrum}
\end{figure*}
We must then set the distance, mass, and spin for each IBH to calculate the radiation flux.
The angle between the observer and each IBH concerning the center in the Galactic plane, $\theta_{\rm BH}$, is given uniformly.
Then, the distance between observer and IBHs is $d=[(R^2+R_{\rm obs}^2-2RR_{\rm obs}\cos\theta_{\rm BH})+z^2]^{1/2}$, where $R_{\rm obs}=8.3\,\rm kpc$ is the observer distance from the Galactic center.
The IBH mass function is set as a power-law shape, $dN/dM\propto M^{-\alpha_{\rm m}}$ ($\alpha_{\rm m}\approx3.5$), which mimics the primary mass distribution obtained from the gravitational wave observation of binary-BH (BBH) mergers \citep[][]{Abbott23}.
Minimum and maximum IBH masses are chosen from their posterior values of parameters in the fitting function, $M_{\rm min}\sim5.0M_\odot,\,M_{\rm max}\sim87M_\odot$.
As for the IBH spin, we use two distinct models of the spin populations based on current observational constraints.
One is the \textit{low spin} model, which is set to reproduce the BBH primary spin probability distribution $P_{\beta}(a)\propto a^{\alpha_{\rm a}-1}(1-a)^{\beta_{\rm a}-1}$ \citep[$\alpha_{\rm a}=2.0,\,\beta_{\rm a}=5.0$;][]{Abbott23}, and peaks around $a\sim0.2$. 
The other is the \textit{high spin} model, reproducing the spin distribution obtained from the spectral analysis of Galactic BH X-ray binaries. 
We fit the spin data of 36 NuSTAR-detected BH X-ray binaries in \citet{Draghis24} with a function $P_{\rm LN}(a)\propto\exp[-(\log(1.0-a)-\mu_{\rm a})^2/(2\sigma_{\rm a}^2)]/[(1.0-a)\sqrt{2\pi\sigma_{\rm a}^2}]$ and obtain $\{\mu_{\rm a},\,\sigma_{\rm a}\}=\{-2.67,\,0.70\}$.
To maintain accountability with the number of IBH samples ($10^5$) much lower than $N_{\rm tot}=10^8$, we repeat the random sampling of $\theta_{\rm BH}$, $M$, and $a$ one-hundred times with a different seed for each BH, and combine all realizations to obtain the final result.

\subsection{Radiation Flux from IBH-MADs}\label{ss:flux}
The next step is to calculate radiation flux from IBH-MADs.
We calculate radiation spectra by the method described in \citet{Kimura20, Kimura21, Kimura21b, Kuze22, Kuze24}.
We assume that the magnetic reconnection in the MADs heats plasma and produces non-thermal particles.
We consider synchrotron emission and Comptonization via thermal electrons, synchrotron emissions from primary electrons and protons, and synchrotron emissions from secondary-produced electrons/positrons via the Bethe-Heitler process and the two-photon annihilation.
\begin{figure*}[t]
 \includegraphics[keepaspectratio, trim = 5 0 0 0, scale=0.45]{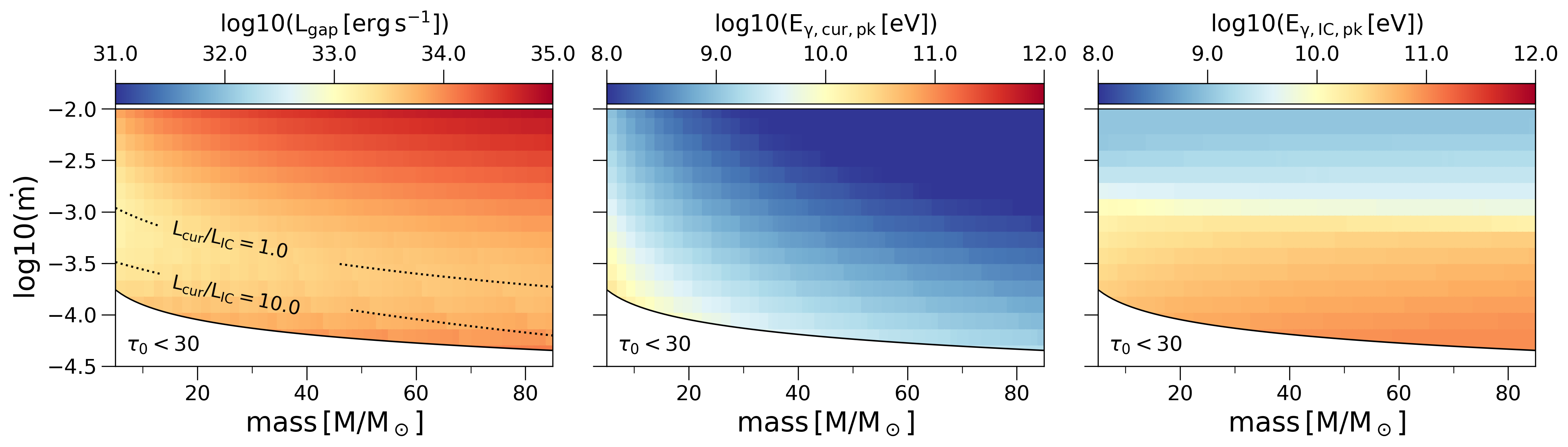}
\caption{Color maps for $L_{\rm gap}$ (left), $E_{\rm \gamma, cur,pk}$ (middle), and $E_{\rm \gamma, IC,pk}$ (right) as functions of $M$ and $\dot{m}$, with $a=0.9$.}
\label{c_gap}
\end{figure*}

The size of MAD is set as $R_{\rm MAD}=\mathcal{R_{\rm MAD}}\rg\,(\mathcal{R_{\rm MAD}}=10)$, where $\rg=GM/c^2$ is the BH gravitational radius.
The magnetic field and the proton number density in the MAD, $B$ and $N_{p}$, respectively, are derived as functions of $M$ and $\dot{M}$ for the fixed viscosity parameter $\alpha\approx0.3$ \citep[cf.][]{Shakura73} and plasma beta $\beta=0.1$.
The fraction of the dissipated energy to the accretion flow gravitational energy is $\epsilon_{\rm dis}=0.15$, and that of the non-thermal particle energy to the dissipated energy is $\epsilon_{\rm NT}=0.33$.
The thermal electrons heating ratio compared to protons is calculated by using the prescription from \citet{R17, Chael18}, $f_{e}\approx0.5\exp{[-(1-4\beta\sigma_{\rm B})/(0.8+\sqrt{\sigma_{\rm B}})]}$, where $\sigma_{\rm B}=B^2/4\pi N_{p}\m c^2\simeq0.5\,\mathcal{R}_{\rm MAD,1}^{-1}\beta_{-1}^{-1}$ is the magnetization parameter.
Then, the electron heating rate can be estimated as
\begin{equation}
\label{qeth}
\begin{aligned}
Q_{e,\rm thermal}
&=f_{e}\epsilon_{\rm dis}(1-\epsilon_{\rm NT})\dot{M}c^2\\
&\simeq3.7\times10^{33}\left(\frac{f_{e}\epsilon_{\rm dis}(1-\epsilon_{\rm NT})}{0.3\cdot0.15\cdot0.67}\right)\dot{m}_{-4}M_1\,\rm{erg\,s^{-1}}.
\end{aligned}
\end{equation}
The electron temperature normalized by the electron rest mass, $\theta_{e}=kT_{e}/\me c^2$ ($k$ is the Boltzmann constant), can be approximated by equating $Q_{e\rm, thermal}$ and the cooling rate via synchrotron radiation, $E_\gamma L_{E_\gamma}$.\footnote{We consider cooling by thermal Comptonization as well, but is sub-dominant for current set of parameters.}
The MADs are mostly optically thick for synchrotron self-absorption by thermal electrons for our current set of parameters.
Then, we can estimate $E_\gamma L_{E_\gamma}$ by using the expression for the Rayleigh-Jeans low-energy tail as
\begin{equation}
\label{lsynpk}
\begin{aligned}
E_\gamma L_{E_\gamma}&\approx E_{\gamma, \rm syn, pk} L_{E_{\gamma,\rm syn, pk}}\\
&\sim E_{\gamma\rm, syn,pk}\cdot4\pi^2R_{\rm MAD}^2\cdot\dfrac{2\me E_{\gamma\rm , syn,pk}^2}{h^3}\theta_{\rm e},
\end{aligned}
\end{equation}
where $E_{\rm\gamma, syn, pk}=x_{\rm M}(3heB\theta_{e}^2/4\pi\me c)$
is the peak energy, $x_{\rm M}$ is the factor representing the deviation of the actual spectral peak from the synchrotron characteristic frequency \citep[see][]{Mahadevan97}, $h$ is the Planck constant, and $e$ is the electric charge.
Therefore, from Eqs.~(\ref{qeth}) and~(\ref{lsynpk}), we obtain $\theta_{e}\simeq1.5\,\dot{m}_{-4}^{-1/14}M_1^{1/14}\mathcal{R}_{\rm MAD,1}^{1/4}\alpha_{-0.5}^{3/14}\beta_{-1}^{3/14}x_{\rm M,2}^{-3/7}$, and the corresponding peak energy is
\begin{equation}\label{synpk}
\begin{aligned}
E_{\rm\gamma, syn, pk}\simeq 2.2\,\dot{m}_{-4}^{5/14}&M_1^{-5/14}\mathcal{R}_{\rm MAD,1}^{-3/4}\\
&\times\alpha_{-0.5}^{-1/14}\beta_{-1}^{-1/14}x_{\rm M,2}^{1/7}\,\mathrm{eV}.
\end{aligned}
\end{equation}
We also introduce $\tau_0=n_0\sT\rg$ ($n_0$ denotes the peak photon number density), which characterizes the radiative compactness for the size of the region in the magnetosphere where MAD thermal photons are dense, $r\lesssim R_{\rm MAD}$ (see Section~\ref{ss:gapdyn}).
Due to the narrow spectral shape of thermal synchrotron photons ($E_\gamma L_{E_\gamma}\propto E_\gamma^3$ for $E_\gamma<E_{\gamma,\rm syn, pk}$ and decaying exponentially comparable to/faster than $\propto E_{\gamma}^{-1}$ above there, see Fig.~\ref{spectrum}), we can evaluate it as
\begin{equation}
\begin{aligned}
\label{t0cal}
\tau_0&=\dfrac{4\pi}{c}\dfrac{E_{\gamma,\rm syn,pk}L_{E_{\gamma,\rm syn,pk}}}{4\pi^2R_{\rm MAD}^2E_{\gamma,\rm syn,pk}}\sT\rg\\
&\simeq50\,\dot{m}_{-4}^{9/14}M_1^{5/14}\mathcal{R}_{\rm MAD,1}^{-5/4}\alpha_{-0.5}^{1/14}\beta_{-1}^{1/14}x_{\rm M,2}^{-1/7}.
\end{aligned}
\end{equation}

The number distributions of non-thermal particle species $i$ are derived from the energy transport equation with one-zone steady-state approximations, including injection, cooling, and escaping processes \citep[Eq.(3) of][]{Kimura20}.
The injection term is given by $\dot{N}_{E_{i,\rm inj}}\propto E_{i}^{-s_{\rm inj}}\exp\left(-E_{i}/E_{i,\rm cut}\right)$, where $E_{i,\rm cut}$ is the cutoff energy.
We use $s_{\rm inj}=2$ based on the long-term 3D PIC simulation \citep{Zhang23}, which is different from that used in \citet{Kimura21}.
The luminosities of non-thermal components are determined to satisfy 
$\int\dot{N}_{E_{p\rm,inj}}E_{p}dE_{p}\approx\epsilon_{\rm dis}\epsilon_{\rm NT}\dot{M}c^2$ for protons and $\int\dot{N}_{E_{e\rm,inj}}E_{e}dE_{e}\approx f_{e}\epsilon_{\rm dis}\epsilon_{\rm NT}\dot{M}c^2$ for electrons, respectively.

In Fig.~\ref{spectrum}, we demonstrate the IBH-MAD spectra for $(M/M_\odot,\,\dot{m})=(10,\,1.0\times10^{-2}),\,(10,\,1.8\times10^{-4}),\,(50,\,6.2\times10^{-5})$.
Thermal synchrotron emission produces the peak around the optical-UV band (Eq.~(\ref{synpk})).
Several humps in around 10-$10^3$ eV are produced by thermal Comptonization.
The primary electrons' non-thermal synchrotron emission dominates the spectrum from keV to MeV.
Synchrotron emission from primary protons and secondary-produced electrons/positrons peak around the MeV-GeV band but are sub-dominant for most of the parameter range of our interest. (Thus not depicted in the figure.)

\subsection{Spark Gap Dynamics \& Relation to MAD Properties}\label{ss:gapdyn}
Here, we evaluate the physical quantities of the magnetosphere and highlight its relation to the MAD.
The magnetic field at the horizon, $\Bh$, is set by the magnetic flux accumulation rate, which depends on $\dot{M}$.
Based on the results from GRMHD simulations \citep[e.g.][]{Tchekho11, Narayan12}, we can evaluate it as $\Bh=\phi\sqrt{\dot{M}c}/2\pi\rg\simeq1.1\times10^7(\phi/50)\dot{m}_{-4}^{1/2}M_1^{-1/2}\,\rm G$, where $\phi$ is the saturated fraction of the normalized magnetic flux.
Inside the magnetosphere, the extraction of the BH rotation energy as the Poynting flux, i.e. the Blandford-Znajek (BZ) process \citep{BZ77} is onset.
The total Poynting luminosity via the BZ process, $\Lbz$, is evaluated as
\begin{equation}\label{lbz}
\begin{aligned}
\Lbz&=\dfrac{\kappa_{\rm B}\pi c}{4}a^2\Bh^2\rg^2\\
&\simeq2.7\times10^{35}\left(\dfrac{\kappa_{\rm B}}{0.053}\right)\left(\dfrac{\phi}{50}\right)^2\left(\dfrac{a}{0.9}\right)^2\dot{m}_{-4}M_1\,\rm erg\,s^{-1},
\end{aligned}
\end{equation}
where $\kappa_{\rm B}=0.053$ is the empirical factor for the split-monopole field configuration \citep{Tchekho10}.
A fraction of this energy input to the magnetosphere will be converted to radiation in the spark gap.

The electron-positron pair production will modify the charge/current distribution in the magnetosphere, thereby controlling the spark gap dynamics and the peak amplitude of the gap electric field.
The pair production by two MeV energy photons emitted from MADs is sub-dominant (see \citealp{Levinson11, Hirotani16, Hirotani18, Kin24}, but also \citealp{Wong21}).
Thus, we mainly consider the reaction between high-energy gamma rays emitted by gap-accelerated particles and MAD thermal synchrotron photons. 
The pair production optical depth for a gamma-ray photon of the energy $E_{\gamma,\rm g}$ in $r\leq R_{\rm MAD}$ can be evaluated by using the thermal synchrotron photon number density, $n_{\rm \gamma, syn}(E_{\gamma,\rm syn})$, and the cross-section, $\sgg(E_{\gamma,\rm syn},\, E_{\gamma\rm,g})$, as 
\begin{equation}
\begin{aligned}
\label{tggmad}
\tggmad(E_{\gamma\rm,g})&= n_{\rm syn}(E_{\gamma,\rm syn})\sgg(E_{\gamma,\rm syn},\,E_{\gamma,\rm g})R_{\rm MAD}\\
&\approx0.1\mathcal{R}_{\rm MAD}\tau_0\cdot\tilde{n}_{\rm syn}(x)\cdot\left(\dfrac{E_{\gamma,\rm syn}E_{\gamma,\rm g}}{\me^2c^4}\right)^{-1}.
\end{aligned}
\end{equation}
In the second line, $n_{\rm syn}$ is normalized as $n_{\rm syn}(E_{\gamma,\rm syn})=n_0\tilde{n}_{\rm syn}(x)$ ($x\equiv E_{\gamma,\rm syn}/E_{\gamma,\rm syn,pk}$), 
and $\sgg$ is substituted by the asymptotic expression, $\sgg(E_{\gamma,\rm syn},\,E_{\gamma,\rm g})\approx0.1\sT[E_{\gamma,\rm syn}E_{\gamma,\rm g}/(\me^2c^4)]^{-1}$ (for $E_{\gamma,\rm syn}E_{\gamma,\rm g}\gtrsim\me^2c^4$).
$E_{\gamma\rm,syn}$ is determined to maximize $\tggmad$, and thus should satisfy the condition $E_{\gamma\rm,syn}=\mathrm{max}\{E_{\gamma,\rm syn,pk},\,E_{\gamma\rm,g}^{-1}\me^2c^4\}$.
By substituting $\tau_0$ obtained from Eq.~(\ref{t0cal}), we can expect $\tggmad\gtrsim1$ for interaction with $E_{\gamma\rm, syn}\sim E_{\gamma\rm,syn,pk}$ soft photons and gamma-ray photons with the energy $E_{\gamma\rm,g}\sim E_{\gamma\rm,syn,pk}^{-1}\me^2c^4\simeq117$ GeV when $M\gtrsim10M_\odot$ and $\dot{m}\gtrsim10^{-4}$, meaning that gamma-ray photons of this energy could produce electron-positron pairs in the magnetosphere.
This is consistent with our GRPIC simulation \citep{Kin24}, in which the spark gap exhibits a steady quasi-periodic oscillation for $\tau_0\geq30$ and bright gamma rays via the curvature process and inverse Compton (IC) up-scattering of MAD optical photons were emitted.
We developed empirical relations between $\tau_0$ and averaged values of various physical quantities during the oscillating state, including maximum electron energy, $\gammae$, and radiation efficiencies for curvature and IC radiation, $\Lcur/\Lbz$ and $\Lic/\Lbz$, respectively: $\gammae\approx8.9\times10^6(\tau_0/30)^{-2/5}$, $\Lcur\approx10^{-2}(\tau_0/30)^{-14/5}\Lbz$, and $\Lic\approx5.8\times10^{-4}\Lbz\propto\tau_0^{-1/5}$.
Those relations are weakly dependent on the shape of the soft photon spectrum and $a$ \citep[see][and also Appendix~\ref{appB} of this paper]{Kin24}, especially for $\tau_0\lesssim100$, for which bright gamma-ray emission is expected.
Therefore, we hereby use those empirical relations for all ranges of parameters to derive the persistent gamma-ray flux.

\subsection{Radiation Flux from IBH Spark Gaps}\label{ss:gapf}
Using $\Lbz$, $\tau_0$, and the empirical relation from our simulation, we calculate the gamma-ray emission spectra from the IBH spark gaps, partially following the method described in Appendix B of \citet{Kin24}.
The characteristic energy of curvature emission from $\gamma_{e}\sim\gammae$ particles in the gap is calculated as
\begin{equation}
\begin{aligned}
E_{\rm\gamma, cur,pk}&=\dfrac{3}{4\pi}\dfrac{hc}{\rg}\gammae^3
\simeq7.7\left(\dfrac{\tau_0}{50}\right)^{-6/5}M_1^{-1}\,\rm GeV,
\end{aligned}
\end{equation}
where the characteristic curvature radius of the magnetic field is assumed to be $\rg$.
You can see that $E_{\rm\gamma, cur, pk}^{-1}\me^2 c^4>E_{\rm\gamma,syn,pk}$.
By substituting a rough approximation $\tilde{n}_{\rm syn}(x)\sim x^{-2}$ (for $x\geq1$, see Fig.~\ref{spectrum}) into Eq.~(\ref{tggmad}), we obtain $\tggmad(E_{\rm\gamma, cur, pk})\simeq0.2(\tau_0/50)^{-7/5}(E_{\rm\gamma, syn, pk}/2.2\,\rm{eV})^2\,M_1^{-2}\mathcal{R}_{\rm MAD,1}$.
Combining with a negative $\dot{m}$ dependence of $E_{\rm\gamma, cur, pk}$ (see Eqs.~(\ref{synpk}) and ~(\ref{t0cal})), we can expect that the most of curvature photons escape from the system.
On the other hand, the energy of photons up-scattered by $\gamma_{e}\sim\gammae$ particles is $\gammae\me c^2\simeq3.7(\tau_0/50)^{-2/5}$ TeV (assuming the reaction in the Klein-Nishina regime), and then $\tggmad(\gammae\me c^2)\sim1.6(\tau_0/50)^{7/5}(E_{\rm\gamma, syn, pk}/2.2\,\rm{eV})^{-1}\,\mathcal{R}_{\rm MAD,1}>1$.
Secondary-produced pairs will repeat a similar interaction sequence until $\tggmad$ becomes $\lesssim1$.
The cutoff energy of IC emission spectra, $E_{\rm\gamma, IC,pk}$, is determined as
\begin{equation}
\begin{aligned}
E_{\rm\gamma, IC,pk}
\simeq17\left(\dfrac{\tau_0}{50}\right)^{-1/2}\left(\dfrac{E_{\rm\gamma, syn,pk}}{2.2\,\rm{eV}}\right)^{-1}\mathcal{R}_{\rm MAD,1}\,\rm{GeV},
\end{aligned}
\end{equation}
where we assumed that $E_{\gamma,\rm syn}\sim E_{\rm\gamma, IC,pk}^{-1}\me^2 c^4$.
In the actual calculation, we use the expression of $\sgg$ in \citet{Coppi90} to obtain $E_{\rm\gamma, IC,pk}$.

Fig.~\ref{c_gap} shows the dependence of $L_{\rm gap}=L_{\rm cur,pk}+L_{\rm IC,pk}$, $E_{\gamma, \rm cur,pk}$, and $E_{\rm \gamma, IC,pk}$ on $M$ and $\dot{m}$ for $a=0.9$.
$L_{\rm gap}$ for $\dot{m}\gtrsim10^{-4}$ is around $10^{33}$ - $10^{35} \,\rm erg\,s^{-1}$.
Both components peak in 1-100 GeV for $\dot{m}\sim10^{-3}$ - $10^{-4}$, meaning GeV-TeV detectable IBHs are in this $\dot{m}$ regime.
The white-filled region shows $\tau_0<30$.
The gap gamma-ray spectra for $a=0.9$ are demonstrated with cyan and magenta dashed lines in Fig.~\ref{spectrum}.
Gap gamma-ray emissions are typically brighter than MAD emission components in GeV-TeV for our current parameter sets.
Thus, we focus on the detectability of gap emission in the GeV-TeV band.

\begin{figure*}[t]
\hspace*{1.0cm}
 \includegraphics[keepaspectratio, trim = 0 0 0 0, scale=0.5]{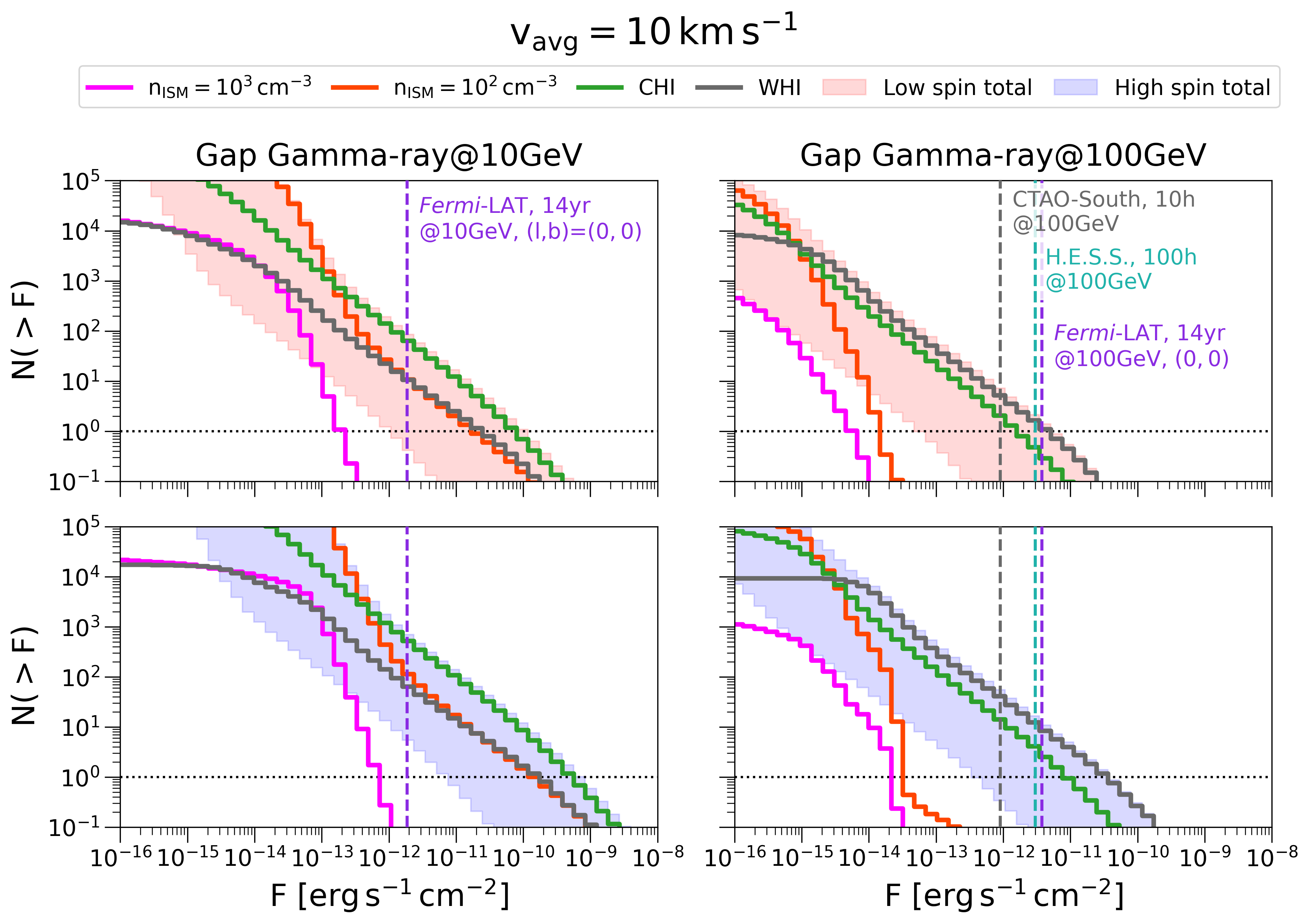}
 \caption{The cumulative distribution function of IBH gamma-ray flux, $\cum$, as functions of gamma-ray flux at 10 GeV (left) and at 100 GeV (right). Red and blue shaded regions denote the sums of all ISM phases for $10^{-2}\leq f_{\rm duty}\leq1.0$, assuming the \textit{low spin} and the \textit{high spin} model, respectively. Contributions of each ISM phase for $f_{\rm duty}=1.0$ are shown by magenta, orange, green, and gray solid lines in all panels.
 The sensitivity limits of \textit{Fermi}-LAT (purple, at 10 GeV and 100 GeV, $(l,\,b)=(0,\,0)$, 14 y; \citet{DR4}), H.E.S.S. (teal, at 100 GeV, 100h; \url{https://www.mpi-hd.mpg.de/HESS/}), and CTAO (grey, South, 50h, at 100 GeV; \url{https://www.cta-observatory.org}) are also shown by thin dashed lines.}
\label{cnum}
\end{figure*}
\begin{figure*}[t]
\hspace*{0.5cm}
 \includegraphics[keepaspectratio, trim = 5 0 0 10, scale=0.57]{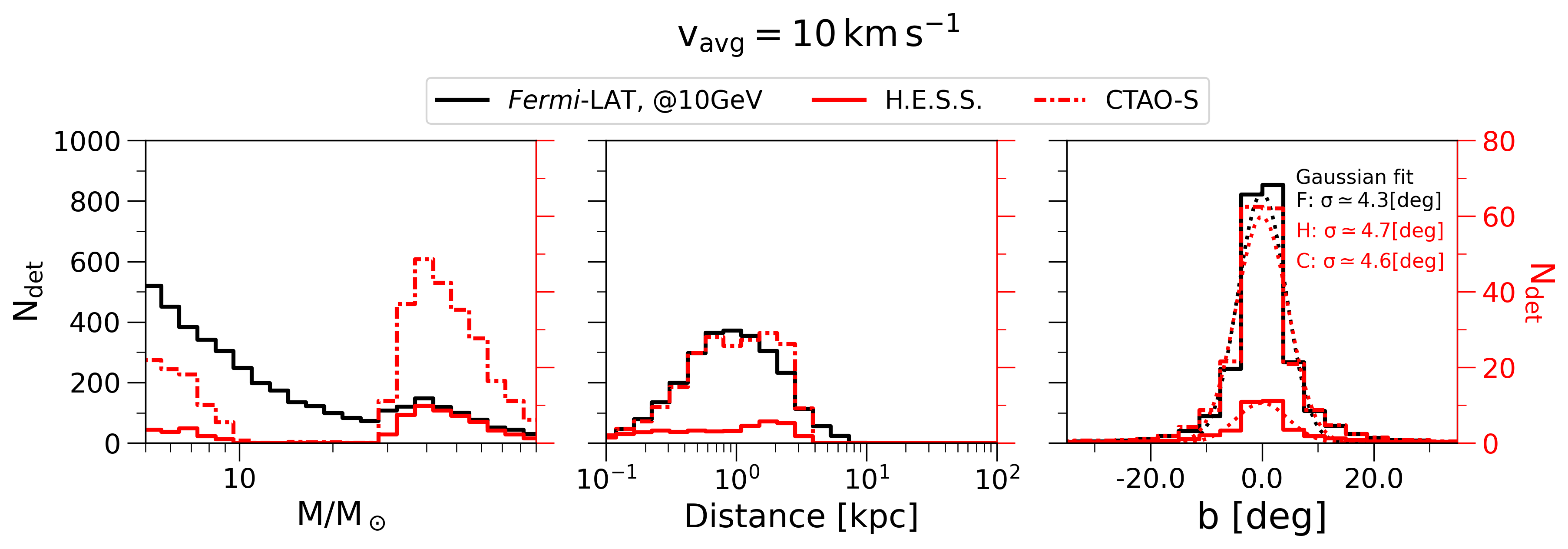}
\caption{Histograms of $M$, $d$, and $b$ for IBHs detectable with \textit{Fermi}-LAT at 10 GeV (black solid), H.E.S.S. at 100 GeV (red solid), and CTAO at 100 GeV (red dash-dotted) for $v_{\rm avg}=10\,\rm km\,s^{-1}$ and $f_{\rm duty}=1$. $\Ndet$ for \textit{Fermi}-LAT in the graph is represented on the left y-axis, while that of H.E.S.S. and CTAO are on the right red y-axis. In the 3rd panel, the results of a Gaussian fitting are shown with thin dotted lines, and the corresponding standard deviation values, $\sigma$, with texts.} 
\label{p_dist}
\end{figure*}

\subsection{Observed Flux Calculation}\label{ss:reduc}
After the consideration above, the intrinsic radiation flux for each IBH is obtained as the function of $M$, $v$, $\nism$, $a$, and $d$, assuming that emissions from both MADs and magnetospheric gaps are isotropic (but see Section~\ref{ss:cav}).
Now, we discuss the attenuation and reduction of gamma-ray fluxes.
Gap gamma-ray photons could interact with protons in the ambient gas via the Bethe-Heitler process, but will be negligible for gas clouds with the column density of $N_{\rm H}\lesssim10^{22}\,\rm cm^{-2}$ due to its low cross section, $\sigma_{\rm BH}\sim10^{-26}\,\rm cm^{2}$.
Attenuation by photons emitted from ambient gas, interstellar radiation field, extragalactic background light, and cosmic microwave background radiation will also be present. Still, it will only be effective at $\gtrsim100$ TeV \citep[e.g.][]{Vernetto16}.
Therefore, we only consider the reduction in the observed gamma-ray flux due to the intermittency of the gap:
the gap opens with a typical duration $t_{\rm gap}\sim1$-$10\rg/c\sim10^{-4}$-$10^{-3}M_1\,\rm s$ during the oscillating state \citep[see][]{Chen20, Kisaka20, Kisaka22, Hirotani22, Kin24}.
Moreover, a luminosity fluctuation induced by the variability in the surrounding accretion flow is expected.
Its typical duration will be the viscous timescale of the accretion flow, $t_{\rm vis}\sim (2\pi/\alpha\Omega_{\rm K})(R_{\rm MAD}/H)^2\sim0.1\alpha_{-0.5}^{\,-1}\mathcal{R}_{\rm MAD,1}^{7/2}(H/0.5R_{\rm MAD})^{-2}M_1 \, \rm s$, where $\Omega_{\rm K}=\sqrt{GM/R_{\rm MAD}^3}$ is the Keplerian angular velocity and $H$ is the disk scale height.
Current gamma-ray detectors, however, would not capture such a short-timescale variability.
Instead, the observed luminosity would be reduced by a factor of the unknown duty cycle.
Our simulation shows the oscillation cycle of $f_{\rm duty}\sim0.1$-$1$ (see Fig.~\ref{simu} in Appendix \ref{appB}).
A naive expectation for the case of the variability produced by the accretion flow would be $f_{\rm duty}\sim t_{\rm gap}/t_{\rm vis}\sim10^{-2}$.
Considering these, we multiply the factor $10^{-2}\leq f_{\rm duty}\leq1.0$ to the intrinsic gamma-ray flux to obtain the observed flux.
We will discuss other possibilities in Section~\ref{ss:cav}.

\section{Detection prospects of gamma-ray emitting IBHs}\label{sec:res}
\begin{figure*}[t]
\includegraphics[keepaspectratio, trim = 0 0 0 0, scale=0.5]{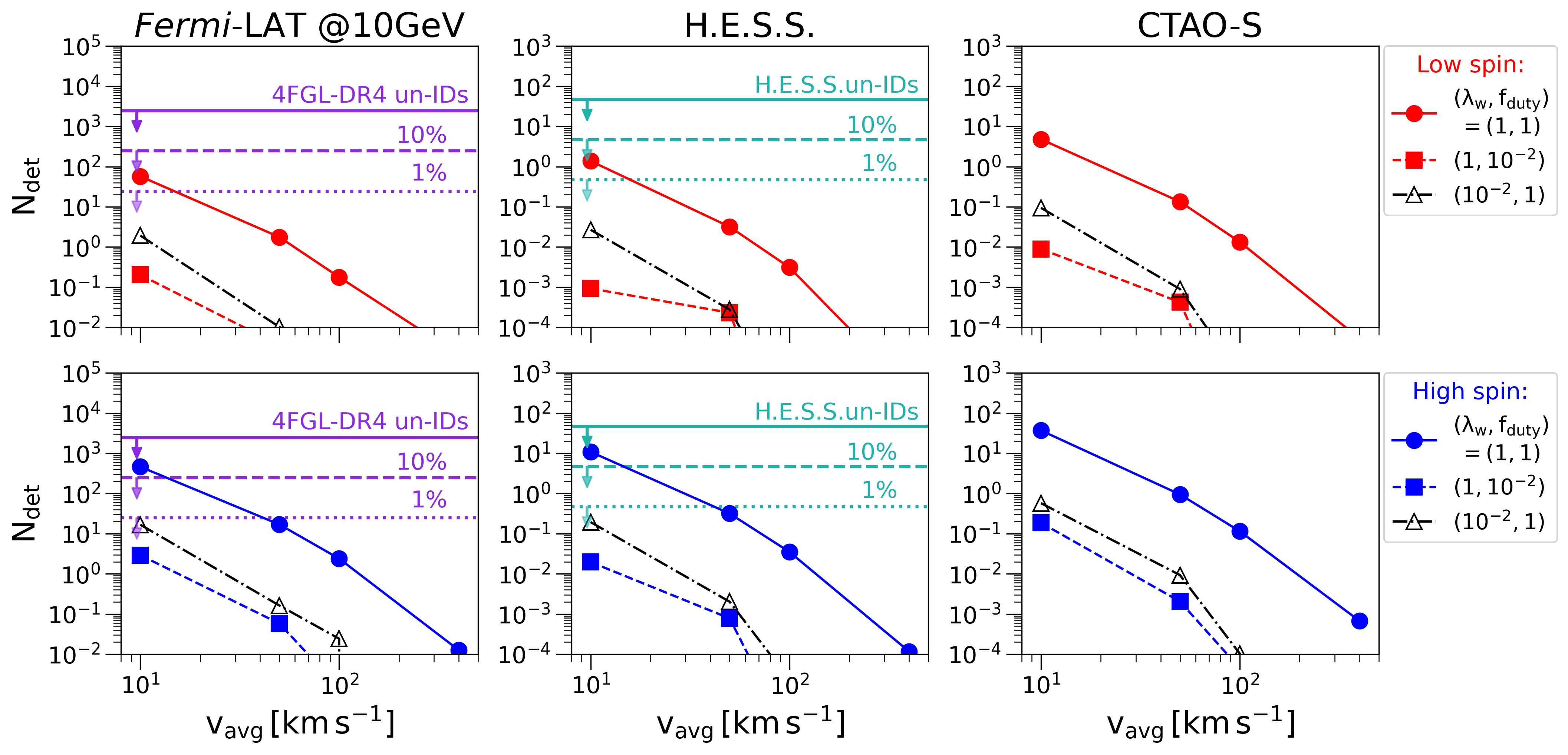}
\caption{$\Ndet$ above the sensitivity limits of \textit{Fermi}-LAT (left), H.E.S.S. (middle), and CTAO (right). The top panels show the results for \textit{low spin} and the bottom panels are for \textit{high spin}. Results for $(\lambda_{w},f_{\rm duty})=(1.0,1.0),\,(1.0,10^{-2}),\,(10^{-2},1.0)$ are shown by solids lines with circles, dashed lines with square, and dash-dotted lines with triangles, respectively.
Purple and teal horizontal solid/dashed/dotted lines in the left and middle panels represent the total number, 10\%, and 1\% of un-IDs in \textit{Fermi}-LAT 4FGL-DR4 \citep{DR4} and H.E.S.S. Galactic plane survey \citep[HGPS;][]{HESS18}.}
\label{cn_v}
\end{figure*}
\subsection{Cumulative number \& Properties for $v_{\rm avg}=10\,\rm km\,s^{-1}$}\label{res1}
In Fig.~\ref{cnum}, we show the cumulative number of expected gamma-ray flux, $\cum$, at $10$ GeV (left) and 100 GeV (right) as a function of gamma-ray flux in the case of $v_{\rm avg}=10\,\rm km\,s^{-1}$.
Here and hereafter, all the results of the detection number are normalized with $N_{\rm tot}=10^8$.
Red and blue shaded regions in the figure represent combined values of $\cum$ for all ISM phases, with $10^{-2}\leq f_{\rm duty}\leq1$.
$\cum$ for \textit{high spin} model will be $\cum\sim10^3$ for \textit{Fermi}-LAT, $\sim10$ for H.E.S.S., and $\sim10^2$ for CTAO at their maximum.
Depending on the value of $f_{\rm duty}$, the detection by all three detectors will be reduced by $\sim10^{-2}$.
The main contribution in all detectors will be IBHs embedded in cold and warm HI gas.
IBHs in dense molecular gas clouds are rather dim, especially in the 100 GeV band.
This is because MAD thermal photons for IBHs in molecular gas clouds are brighter and have peaks in higher frequency than in cold/warm HI gas due to a higher $\dot{M}$, leading to a stronger attenuation in $\sim$ 100 GeV.

Next, in Fig.~\ref{p_dist} we show the number histograms of $M$, $d$, and the galactic latitude $b$ for IBHs with gamma-ray flux above the sensitivity limit of \textit{Fermi}-LAT at 10 GeV ($F\approx10^{-12}\,\rm erg\,s^{-1}\,cm^{-2}$), H.E.S.S. ($F\approx3.0\times10^{-12}\,\rm erg\,s^{-1}\,cm^{-2}$) at 100 GeV, and CTAO ($F\approx10^{-12}\,\rm erg\,s^{-1}\,cm^{-2}$) at 100 GeV.
The mass distributions of detectable IBHs have a bimodal shape peaking at $M \sim5M_\odot$ and $\sim40M_\odot$.
Those peaks correspond to the preferred mass ranges to have a spectral peak at $\sim100$ GeV, which are determined by a negative (positive) dependence of $E_{\gamma,\rm cur, pk}$ ($E_{\gamma,\rm IC, pk}$) on $M$ (see Fig.~\ref{c_gap}).
The reason for the relatively small $M\sim40M_\odot$ peak for \textit{Fermi}-LAT is that the IC component peaking around 100 GeV is more significant than the curvature component for the majority of $M\sim40M_\odot$ IBHs, and then 10 GeV flux will be reduced due to the hard spectral shape.
The distance distribution peaks around $d\sim1.0\,\rm kpc$, which corresponds to the detection horizon for typical IBHs with the typical gamma-ray luminosity $\sim10^{33}\,\rm erg\,s^{-1}$.
The Galactic latitude of each detectable IBH will mainly be in $|b|\lesssim5$ deg, which is understood as the latitude for BHs at the typical distance $d\sim1$ kpc and within the scale height of cold HI, the main contributor to the detection.
This concentration to the Galactic plane is roughly consistent with that of detected un-IDs, showing a quite narrow concentration to the Galactic plane \citep[see][]{HESS18, 4FGL-DR3, DR4}.

\subsection{Dependence on $v_{\rm avg}$, $f_{\rm duty}$, and $\lambda_{w}$}
Fig.~\ref{cn_v} shows the detectable IBH number above the sensitivity limit, $\Ndet$, for \textit{Fermi}-LAT at 10 GeV, and H.E.S.S., CTAO at 100 GeV for various $v_{\rm avg}$.
For \textit{high spin} model, $\Ndet>1$ can be achieved for all three detectors when $\vavg\leq50\,\rm km\,s^{-1}$ and  $f_{\rm duty}\sim1$.
For \textit{low spin} model, on the other hand, the number is $10^{-1}$ lower for all $\vavg$, and $\Ndet>1$ by H.E.S.S. can only be achieved when $\vavg\sim10\,\rm km\,s^{-1}$ and $f_{\rm duty}\sim1$.
One reason for an overall negative dependence of $\Ndet$ on $\vavg$ is that more IBHs are within the ISM scale height for the spatial distribution concentrating on the Galactic plane, which can be achieved for lower $\vavg$ (see Fig.~\ref{dist} in Appendix \ref{appA}).
Also, $v\lesssim a\,few\times10\,\rm km\,s^{-1}$ is preferable to achieve $\tau_0>30$ (see Eqs.~(\ref{acc_rate}) and (\ref{t0cal})).

We show the results for $\lambda_{w}=10^{-2}$ and $f_{\rm duty}=1$ by black dash-dotted lines with triangles in Fig.~\ref{cn_v}.
Compared to the $\lambda_{w}=1$ case, $\Ndet$ is reduced by $10^{-2}$ for $v_{\rm avg}\lesssim10^2\,\rm km\,s^{-1}$ and goes to zero for higher $\vavg$.
We can still expect $\Ndet\sim1$-$10$ for \textit{high spin} and $f_{\rm duty}=1$ by \textit{Fermi}-LAT, but $\sim0.1$ by H.E.S.S. and CTAO.

\section{Discussions}\label{sec:disc}

\begin{figure}[t]
\hspace*{0.2cm}
\includegraphics[keepaspectratio, trim = 0 0 0 0, scale=0.48]{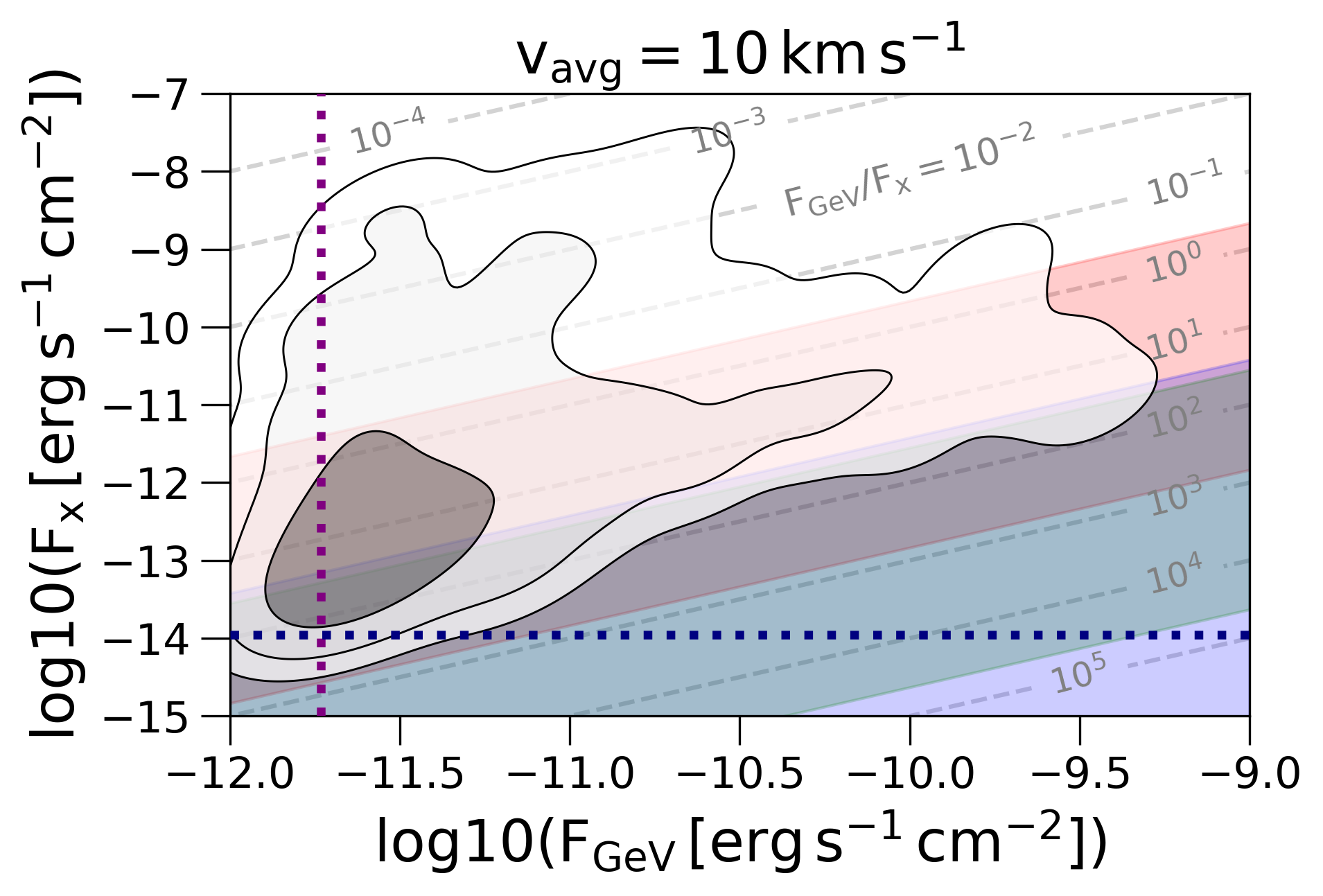}
\caption{the $\sigma$ (black), $2\sigma$ (gray), and $3\sigma$ (silver) regions that IBHs both detectable by eROSITA and \textit{Fermi}-LAT at 10 GeV occupy in $F_{\rm GeV}$-$F_{\rm X}$ plane for the \textit{low spin} case, $\vavg=10\,\rm km\,s^{-1}$.
Differential sensitivity limits for the two detectors are shown by navy and purple dotted lines, respectively.
$2\sigma$ filled region of $F_{\rm GeV}/F_{\rm X}$ for blazers (red), milli-second pulsars (green), and young pulsars (blue) both detected by \textit{Fermi}-LAT and eROSITA are shown for comparison \citep[see][]{Mayer24}.}
\label{fgx}
\end{figure}

\subsection{Strategy for Detection}\label{strate}
Our IBH candidates will be bright in optical, X-ray, and 10-100 GeV bands.
A main strategy for the detection is thus taking cross-correlations of all-sky survey catalogs in those three bands.
\citet{Kimura21} argued that IBH-MADs occupy similar positions to white dwarfs (WDs) in the color-magnitude (HR) diagram and the optical-X luminosity plane.
By comparison, here we consider a broader range of $v$  than \citet{Kimura21}, including $v\lesssim10\,\rm km\,s^{-1}$. We find that, for $\dot{m}\gtrsim10^{-4}$, where bright gap gamma rays are emitted, intrinsic optical and X-ray luminosities will be $\times10$ brighter and occupy a slightly bluer regime than typical WDs in the HR diagram.
Dust extinction and absorption by ambient molecular gas must be considered.
However, detectable ones are mainly in cold or warm HI gas, whose gas surface density is low, and hence extinction/absorption is insignificant.

Fig.~\ref{fgx} shows the regime that IBHs detectable in both X-ray and GeV gamma rays occupy in the $F_{\rm GeV}$-$F_{\rm X}$ plane.
As seen from the contour, IBHs with $F_{\rm GeV}/F_{\rm X}\sim1$-$10^2$ are most abundant, which is a much lower ratio than that of pulsars (blue and green).
Differentiation from blazers (red), which have a similar $F_{\rm GeV}/F_{\rm X}$, can be done by parallax observation with \textit{Gaia}, though careful selection of candidates will be required due to the low angular resolution of \textit{Fermi}-LAT.

The IBH X-ray and gamma-ray flux will exhibit time-variability due to the $\dot{M}$ fluctuation.
As for the X-ray band, the flux variation with a timescale $t_{\rm vis}\sim 0.1\alpha_{-1}^{\,-1}\mathcal{R}_{\rm MAD}^{7/2}(H/0.5R_{\rm MAD})^{-2}M_1 \, \rm s$ will be observed for nearby ones.
This would be the key to distinguishing IBHs from X-ray-bright cataclysmic variables, which would have a much longer variability timescale \citep[$\gtrsim10$ s, see e.g.,][]{Nishino22}.
For gamma rays, only $\gg10^3$ s timescale of the variability is detectable for the typical gamma-ray flux $F\sim10^{-11}\,\rm erg\,s^{-1}\,cm^{-2}$.
Such long duration of variability will be produced by the $\dot{M}$ fluctuation with the accretion timescale from the Bondi radius, $r_{\rm B}\sim GM/v^2\sim5.3\times10^{13}M_1(v/50\,\rm km\,s^{-1})^{-2}\,\rm cm$, $t_{\rm acc}\sim r_{\rm B}/v\simeq1.1\times10^7M_1(v/50\,\rm km\,s^{-1})^{-3}\,s$.
The amplitude of luminosity variation will be large enough to detect ($\times10^{\pm2}$) due to the sensitive dependence of $L_{\rm gap}$ on $\tau_0$ and the magnetospheric current amplitude \citep[see Eq.~(\ref{t0cal}), and also][]{Kisaka22, Kin24}.
Multi-epoch observations with CTAO (or \textit{Fermi}-LAT for nearest ones) can capture such variability.

During the interaction of IBHs and the dense ISM, a wake or filament of compressed medium with the scale of $\sim$0.1 pc might be left behind \citep[][]{Nomura18, Kitajima23}.
Morphology analysis of ambient clouds via CO line / 21 cm line might provide another implication to the candidates' location. 

A fraction of accelerated protons diffuse out of the MADs, which would interact with ambient gas to produce ultrahigh-energy gamma rays.
In the case of IBHs embedded in dense molecular clouds or large-sized ($\gtrsim100$ pc) cold HI gas clouds, incident protons from MADs could have energy as high as $\sim$ PeV, and the $pp$ interaction rate could be high due to a high surface density of gas.
Those will produce another spectral peak in the 10-100 TeV band, which could explain some of LHAASO un-IDs \citep[see][]{Kimura24}.

\subsection{Implication to BH physical parameters}\label{imply}
\begin{figure}[t]
\hspace*{0.2cm}
 \includegraphics[keepaspectratio, trim = 0 0 0 0, scale=0.52]{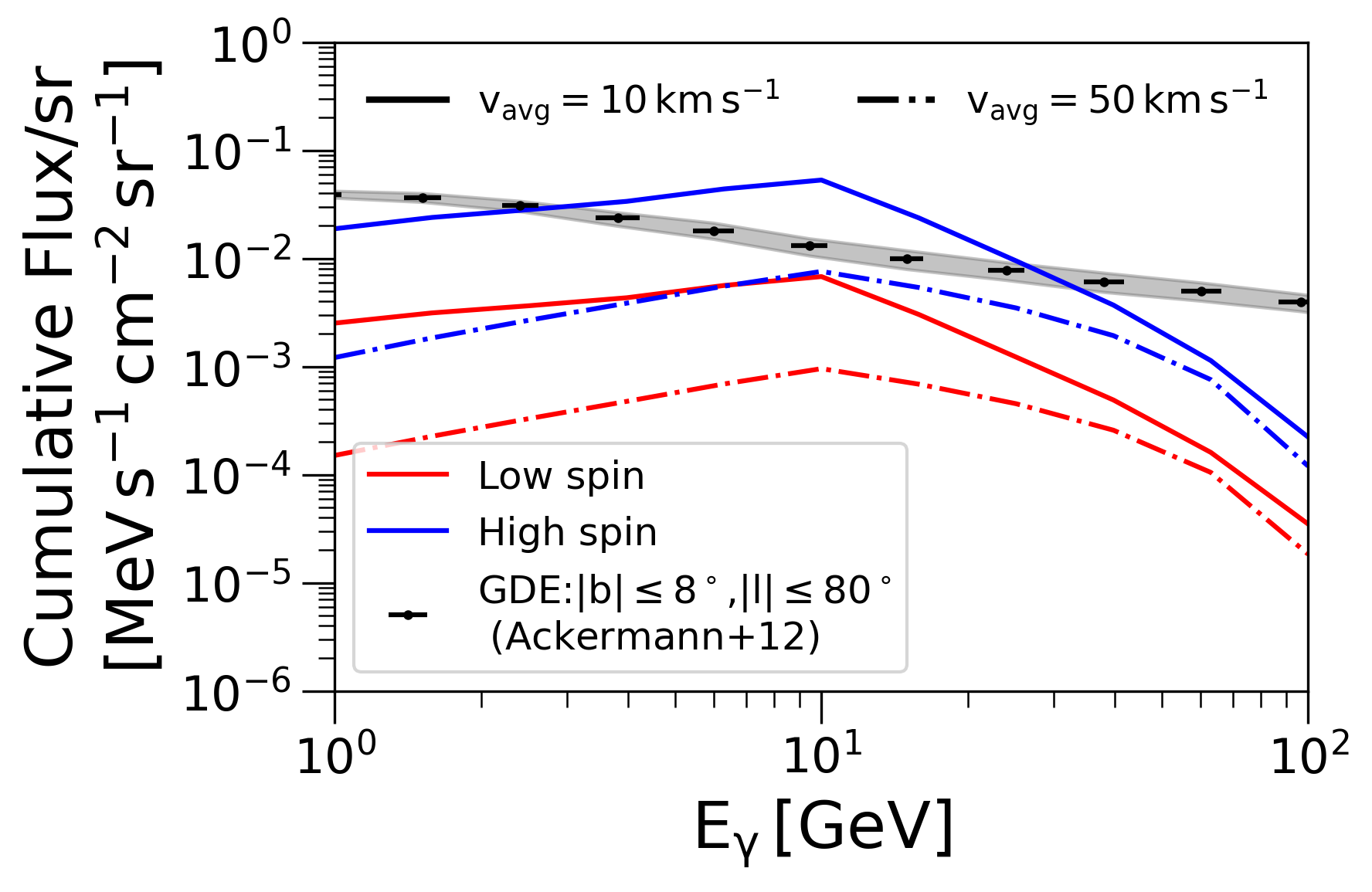}
\caption{Plot of the combined IBH intensity spectra in 1-100 GeV for $\vavg=10\,\rm km\,s^{-1}$, $f_{\rm duty}=1.0$ (solid lines) and $50\,\rm km\,s^{-1}$, $f_{\rm duty}=1.0$ (dash-dot lines).}
\label{cf}
\end{figure}
The thermal synchrotron peak luminosity for gamma-ray detectable IBHs is $L_{\rm opt}\sim10^{34}$-$10^{36}\,\rm erg\,s^{-1}$. 
The peak energy, $E_{\gamma,\rm syn,pk}$, tends to be at a few times higher energy than that of the \textit{Gaia} band, and hence the \textit{Gaia} band luminosity will be $\times10^{-1}$ lower.
And yet, considering their characteristic distance $d\sim1$ kpc (Fig.~\ref{p_dist}), the expected \textit{Gaia} magnitude will be $\sim12-14$.
\textit{Gaia} will be able to major the parallax for such a magnitude, as long as they are in cold/warm HI gas.
Some constraints on their proper motion velocity and BH mass can be obtained from the positions and movements of IBH candidates.
The multi-band spectral fit and analysis of the luminosity variation would give us further information on $M$ and $\dot{M}$.
Also, gap gamma-ray emission will trace the dynamical behavior of the magnetosphere.
Detailed analysis of the emission properties may provide observational clues to unanswered questions on the plasma dynamics in the magnetosphere.

As seen in the leftmost panel of Fig.~\ref{p_dist}, H.E.S.S. and CTAO preferably detect massive ($M\gtrsim30M_\odot$) BHs.
So far, a few tens of BHs with $M\sim30M_\odot$ have been found through the gravitational wave observation, and also \textit{Gaia} astrometrical analysis (Gaia BH3).
Their origins remain unclear, though several possibilities are proposed including very massive stars formed in zero/low-metal environment \citep[e.g.,][]{Kinugawa14, Tanikawa22} and hierarchical mergers in dense star clusters \citep[e.g.,][]{Kumamoto20, Tagawa20}.
If we find several IBH candidates using gamma-ray detectors, estimates of the mass and dynamical motion of IBH candidates might achieve important suggestions on their formation channel.

The \textit{high spin} model with $ \vavg\lesssim10^2\,\rm km\,s^{-1}$ predicts $\Ndet$ comparable to the $\sim10\%$ of \textit{Fermi}-LAT un-IDs (see Fig.~\ref{cn_v}).
Considering the rarity of hard spectral objects in \textit{Fermi}-LAT un-IDs \citep{4FGL-DR3, Yang23, Zhu24}, our results may suggest that $\vavg>10\,\rm km\,s^{-1}$ and low spin values are preferred not to overshoot the current un-ID number.

The contribution of IBHs below the detection limit to the Galactic gamma-ray diffuse emission (GDE) would be non-negligible.
We demonstrate 1-100 GeV combined photon intensities for several parameters in Fig.~\ref{cf}.
Here, we combine BHs inside $|b|\leq8^\circ$ and $|l|\leq80^\circ$ with their radiation flux below the sensitivity threshold of \textit{Fermi}-LAT, and divide them with corresponding area $\Pi\sim0.78\,\rm sr$.
All models peak at $\sim10$ GeV, reflecting the characteristic peak of individual spectra.
We obtain the intensity reaching a fraction ($\sim50\%$ at most) of GDE flux for $\vavg=50\,\rm km\,s^{-1}$.
The intensity for \textit{high spin} and $\vavg=10\,\rm km\,s^{-1}$ case with $f_{\rm duty}=1$ exceeds the observed GDE intensity, which might also put a constraint on the spin value and $\vavg$.
\citet{Koshimoto24} discussed influence of $\vavg$ on the detectability of microlensing events induced by IBHs.
They concluded that $\vavg\lesssim10^2\,\rm km\,s^{-1}$ is preferred to explain the detection of OGLE-2011-BLG-0462, a singular example of the IBH microlensing event.
Combining with our results, $\vavg\sim a\,few\times10\,\rm km\,s^{-1}$ might be a suitable value for the average kick velocity, which is also consistent with the best-fit transverse motion velocity of the IBH responsible for OGLE-2011-BLG-0462 event in \citet[$v\sim37.6\,\rm km\,s^{-1}$;][]{Lam23}.

Above discussion depends on the assumption of $\lambda_{w}$ and $f_{\rm duty}$: we obtain $\times10^{-2}$ lower $\Ndet$ for $\lambda_{w}=10^{-2}$ or $f_{\rm duty}=10^{-2}$, regardless of $\vavg$ and the distribution of $a$ (Fig.~\ref{cn_v}).
In this case, we hardly achieve any constraints on $\vavg$ and $a$ from the detection number or their contribution to GDE.
Recent 3D GRMHD simulations of BHL accretion show $\lambda_{w}\sim0.1$-$0.5$ \citep{Kaaz23, Galishnikova24, Kim24}, and thus our fiducial value $\lambda_{w}=1$ might overestimate $\dot{M}$.
About $10^{-1}$ reduction in $\Ndet$ is expected for $\lambda_{w}\sim0.1$.
\textit{Fermi}-LAT can still detect $\sim10$-$100$ for $\vavg=10\,\rm km\,s^{-1}$ and \textit{high spin}, which corresponds to $\sim1$-$10$\% of un-IDs.

\subsection{Caveats}\label{ss:cav}
Previous attempts to estimate $N_{\rm tot}$ have yielded a variety of values ranging $10^8$-$10^9$, depending on the model for the star-formation history and the stellar evolution \citep[e.g.][]{vanden92, Brown94, Timmes96, Caputo17, 
Olejak20}.
The detectable IBH number is proportional to $N_{\rm tot}$, and then we would expect $\times10$ higher $\Ndet$ for all parameters with $N_{\rm tot}=10^9$, which would give tighter constraints on $\vavg$ and $a$, as discussed in Section \ref{imply}.
The IBH mass function remains uncertain as well.
For a harder (softer) power-law index or a higher (lower) maximum mass than our assumptions, the higher (lower) $\Ndet$ is expected.
Its effect would be more eminent in observations at $\sim100$ GeV than at $\sim10$ GeV, 
since those detect more $M>10M_\odot$ IBHs (see Fig.~\ref{p_dist}).
On the contrary, if the minimum mass is lower (higher) than our assumption, $\Ndet$ at $\sim$ 10 GeV would increase (decrease) because the higher the mass is, the lower $E_{\gamma,\rm cur,pk}$ becomes, and vice versa.

Some uncertainties in the gap dynamics must be noted:
One is the position of the gap.
We have assumed the spherical distribution of the gap around the BH and isotropic gamma-ray emission.
The attenuation by MAD soft photons and Thomson scatterings of accreting plasma is insignificant for 10-100 GeV gap main emission due to low photon/particle number density.
Thus, the radiated photon could escape from the system all over the direction.
In several BH magnetosphere 1D GRPIC simulations \citep[][]{Kisaka20, Chen20, Kisaka22, Kin24}, a spark gap opens around the null surface, which is almost spherical for the split-monopole field geometry \citep[see][]{Hirotani16}.
The monopole-like shape of the magnetic field near the horizon is observed in GRMHD simulations \citep[see][]{Tchekho10, Ripperda22}, and therefore the sphericity of the gap might be valid, leading to the isotropic gamma-ray emissions.
In contrast, recent 2D GRPIC simulations of the BH magnetosphere \citep[e.g.][]{Crin20} observed the gap opening mainly around the polar region.
If the gap opening angle is limited, our expectation of the detection number will be reduced.
Such qualitative differences in behaviors of the magnetosphere between local 1D and global 2D simulations need to be addressed in the future.

The gap duty cycle is also worth investigating because, as shown above, the reduction in the persistent gamma-ray luminosity significantly affects the number of detections.
Other than two characteristic timescales that we mention in Section~\ref{ss:reduc}, the sporadic magnetic reconnection in the equatorial plane will change the global structure of the electric current in the magnetosphere, which might produce a cyclic behavior on the gap activity \citep[see][]{Vos24}.
The pair cascading might also be induced by this sudden magnetic energy release \citep{Kimura22, Chen23, Hakobyan23}.
The amount of injected pair could be large enough to screen the entire magnetosphere, but due to a much longer interval between the flare \citep[$t_{\rm int}\sim10^3\rg/c$, see][]{Ripperda22} compared to the approximated escape timescale of the injected plasma, $t_{\rm esc}\sim 3R_{\rm MAD}/c\sim30\rg/c$, the gap might become active again in between the flares.
A simple estimate for the duty cycle will be $f_{\rm duty}\sim(t_{\rm int}-t_{\rm esc})/t_{\rm int}\sim1$, but a detailed simulation focusing on such the concordance between the magnetosphere and the accretion flow will be required to investigate the precise behavior.

\section{Conclusions}\label{sec:conc}
In this work, we have investigated the prospect of detecting IBHs through high-energy gamma rays from magnetospheric gaps.
We calculate the spatial distribution and the proper motion velocity of IBH samples following the method described in \citet{Tsuna18, Tsuna19}.
We then evaluated the mass accretion rate of each IBH and calculated multi-wavelength radiation flux from their MADs.
Gap gamma-ray flux can be calculated using radiation properties of optical soft photons from the MAD and the dependence of $\gamma_{\rm e, max},\,\Lcur,\,$and $\Lic$ on $\tau_0$, which obtained from our simulations.
Gap gamma rays can be bright if IBHs are embedded in relatively dense ISM, for which $\dot{m}\gtrsim10^{-4}$.
In the case of $\vavg=10\,\rm km\,s^{-1}$, the \textit{Fermi}-LAT un-IDs would contain  $\sim10^3$ of gamma-ray emitting IBHs, and for H.E.S.S., there would be $\sim10$ objects near the Galactic plane, at maximum (see Fig.~\ref{cn_v}).
Future observation by CTAO would detect $\lesssim10^2$ candidates near the Galactic plane, mainly in cold/warm HI gas.
Optical and X-ray emissions from surrounding MADs are expected as counterparts.
A strong time variation in the gamma-ray luminosity is expected, due to a fluctuation in the mass accretion rate, which can be detected through the observation by \textit{Fermi}-LAT or CTAO with $\sim10^3$s operation time.
Those IBHs could also contribute to the GDE, mainly in $\sim$ GeV.
Constraints on the supernova kick velocity and the BH spin could be achieved by comparison with the number of existing un-IDs or observed GDE intensity.

\begin{acknowledgments}
Numerical computations were carried out on Cray XC50, PC cluster, Small Parallel Computers, and analysis servers at the Center for Computational Astrophysics, National Astronomical Observatory of Japan, and Yukawa-21 at Yukawa Institute for Theoretical Physics at Kyoto University.
K.K. thanks Ryunosuke Maeda for practical advice on the computation in this work. 
This work was supported by the Graduate Program on Physics for the Universe (GP-PU), Tohoku University (K.K.), JST SPRING, Grant Number JPMJSP2114 (K.K., R.K.), KAKENHI No. 22K14028 (S.S.K.), 21H04487 (S.S.K.), 23H04899 (S.S.K).
S.S.K. acknowledges the support of the Tohoku Initiative for Fostering Global Researchers for Interdisciplinary Sciences (TI-FRIS) of MEXT’s Strategic Professional Development Program for Young Researchers.
\end{acknowledgments}

\appendix
\section{IBH dynamical calculation results}\label{appA}
Our IBH dynamical calculation follows the methods described in \citet{Tsuna18, Tsuna19}.
The total number of BH particles is $10^5$, which is $\times10$ smaller than in \citet{Tsuna18}.
We set the numerical integration time-step as $\Delta t=10^3\,\rm yr$ (equivalent to $10^7$ divisions of 10 Gyr) for BHs at $r>10$pc, which is shorter/comparable to their dynamical timescale.
For BHs at $r\leq10$pc, we use the $1/100$ shorter time step $\Delta t=10$yr to achieve enough time resolutions.
With our fixed time step treatment, some of IBHs could have artificially high velocity.
To reduce the number of such IBHs, we introduce the smoothing terms to the equation of motion of BHs (Eqs. (9) and (10) of \citet{Tsuna18}):
\begin{equation}\label{EoM}
\begin{aligned}
&\begin{aligned}
\dfrac{dv_{R}}{dt}=&\dfrac{v_{\theta}^2}{\sqrt{R^2+\epsilon_1^2}}-\sum_{i=1,2}\dfrac{GM_iR}{\left\{R^2+[a_i+(z^2+b_i^2)^{1/2}]^2+\epsilon_2^2\right\}^{3/2}}-\dfrac{G\Mh}{\Rh}\dfrac{R}{\sqrt{R^2+\epsilon_i^2}}\left[\dfrac{\Rh}{\sqrt{R^2+\epsilon_i^2}\sqrt{R^2+\Rh^2+\epsilon_2^2}}\right],
\end{aligned}\\
&\begin{aligned}
\dfrac{dv_{z}}{dt}=&-\sum_{i=1,2}\dfrac{GM_iz[a_i+\sqrt{z^2+b_i^2}]}{\left\{R^2+[a_i+\sqrt{z^2+b_i^2}]^2+\epsilon_i^2\right\}^{3/2}\sqrt{z^2+b_i^2}}-\dfrac{G\Mh}{\Rh}\dfrac{R}{\sqrt{R^2+\epsilon_i^2}}\left[\dfrac{\Rh}{\sqrt{R^2+\epsilon_i^2}\sqrt{R^2+\Rh^2+\epsilon_i^2}}\right],
\end{aligned}
\end{aligned}
\end{equation}
where $v_R,\,v_{\theta},\,v_z$ is the $R,\,\theta,\,z$ velocity components of BH particles.
$M_i=\{4.07\times10^9M_\odot,\,6.58\times10^{10}M_\odot\}$, $a_i=\{0,\,4.85\,\rm kpc\}$, $b_i=\{0.184\,\rm kpc,\,0.305\,\rm kpc\}$ ($i=1,\,2$), $\Mh=1.62\times10^{12}M_\odot$, and $\Rh=200\,\rm kpc$ is the values of the constants obtained from \citet{Irrgang13}.
We set $(\epsilon_1,\,\epsilon_2)=(0.1\rm\, pc,\,1.0\,pc)$ when BHs are at $r=(R^2+z^2)^{1/2}\leq10\,\rm{pc}$ during the calculation and $(0.1\rm \,pc,\,10\,pc)$ when they are at $r>10\,\rm{pc}$.
We have checked the number of BHs having the unphysical velocity $v=(v_R^2+v_\theta^2+v_z^2)^{1/2}> v_{\rm esc}(r=10\,\rm pc)\simeq700\rm\,km\,s^{-1}$ are fewer than 0.1\% of the total for such $\epsilon_1,\,\epsilon_2$, which is negligible in overall detectability calculations.
\begin{figure*}[t]
\hspace*{-0.2cm}
 \includegraphics[keepaspectratio, trim = 5 0 0 0, scale=0.4]{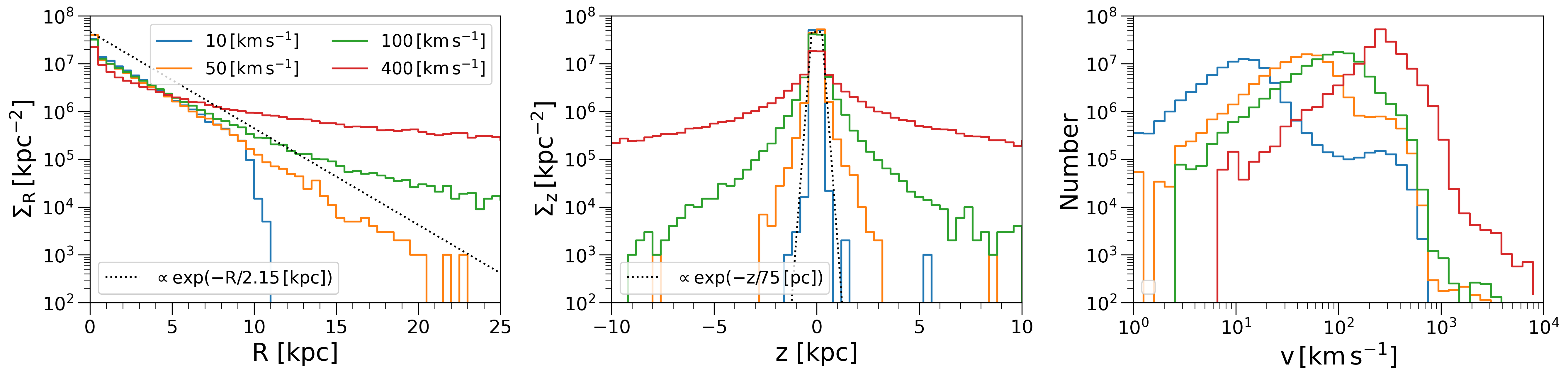}
\caption{Dynamical calculation results of the surface density in $R$ direction (left), the surface density in $z$ direction (middle), and the velocity distribution (right). The black dotted line in each panel represents the initial spatial/velocity distributions.}
\label{dist}
\end{figure*}

We show results of IBH dynamical calculations for 4 different values of $v_{\rm avg}$ in Fig.~\ref{dist}.
Their spatial distribution approaches from the initial exponential disk shape to a more spherical shape for higher $v_{\rm avg}$, consistent with \citet{Tsuna18}.
The velocity distribution peaks at $v\sim v_{\rm avg}$.
The absence of low velocity ($v\lesssim 10\,\rm km\,s^{-1}$) BHs in $\vavg\geq100\,\rm km\,s^{-1}$ is the consequence of the limited number of IBH particles, which might reduce the resultant number of detectable IBHs.

\section{The Summary of Supplemental Analysis of the Spark Gap Dynamical Dependence on the BH Spin}\label{appB}
We hereby summarize our supplemental analysis of the dependence of the spark gap dynamics on the BH spin.
We use the 1D GRPIC simulation code \texttt{Zeltron} \citep{Levinson18}.
We choose the same parameter sets for the system as \citet{Kin24}: BH mass $M=10M_\odot$, the magnetic field at the horizon $\Bh=2\pi\times10^7\,\rm G$, $\theta=30^\circ$, and the characteristic energy of the soft photon fields $\emin=10^{-6}$.
We change the dimensionless spin parameter $a=0.1,\,0.5$ and compare the results with $a=0.9$ runs on \citet{Kin24}.
For each spin value, we perform the run with three different values of fiducial optical depth $\tau_0=30,\,100,\,300$ to investigate the dynamical difference quantitatively.
The initial number of the particles per cell (PPC) is set as 45, which is the same value for \citet{Kisaka22, Kin24}.
We also conducted runs for $a=0.1,\,0.5$ and $\tau_0=30$ with a 3 times higher PPC, 135 (not in the figure), and checked that they reached their convergence.

\begin{figure*}[t]
\hspace*{1.5cm}
 \includegraphics[keepaspectratio, trim = 0 0 0 0, scale=0.55]{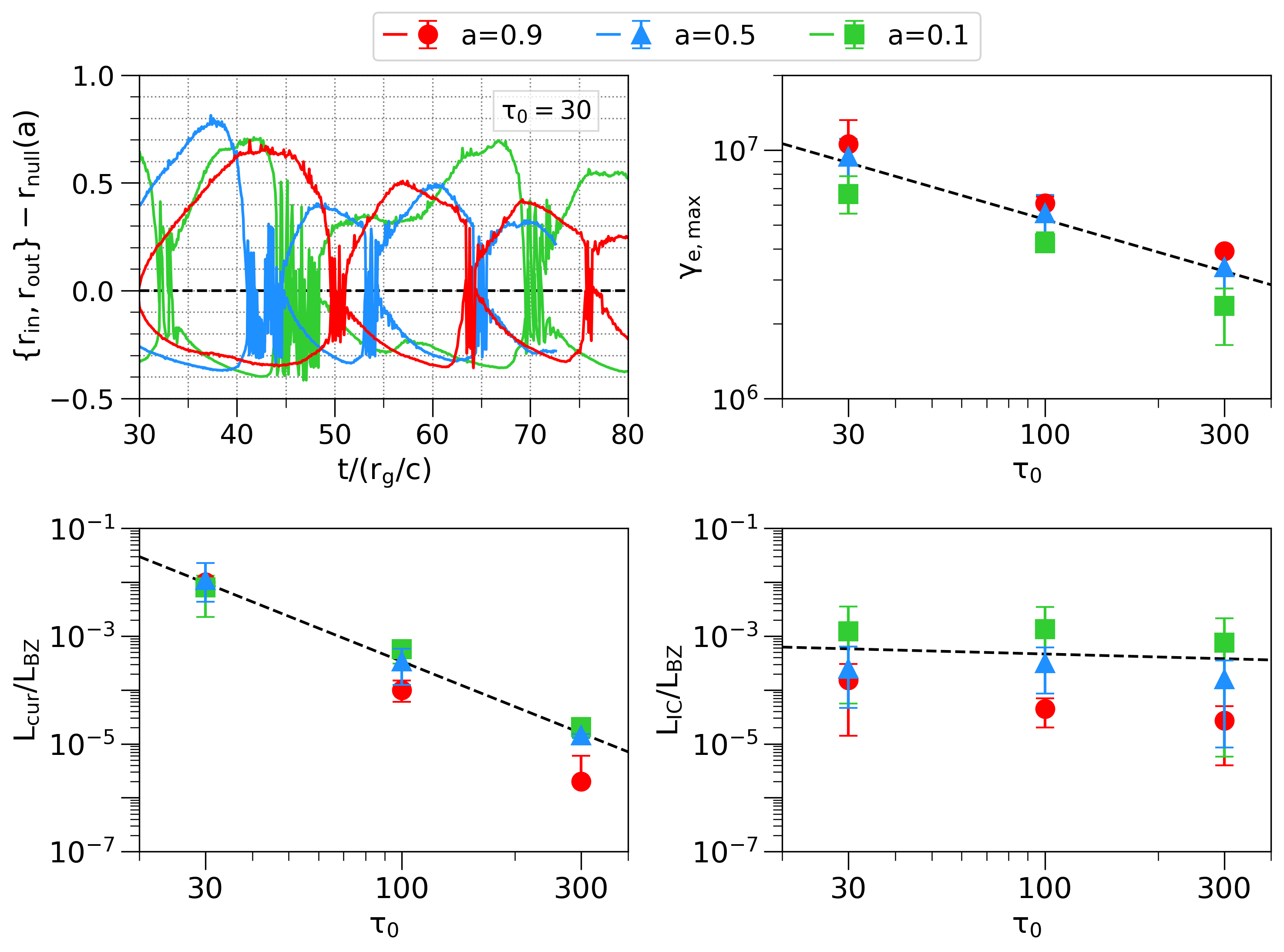}
\caption{Results from our simulation. The results for $a=0.9$ \citep['LA1', 'LA3', and 'LA5' model in][]{Kin24} are shown with red curves/dots for comparison. Each panel shows (top left) the time evolution of spark gap inner/outer radius $r_{\rm in/out}$ around the null $r_{\rm null}$ for $\tau_0=30$, $\gammae$ (top right), $\Lcur$ (bottom left), and $\Lic$ (bottom right) as functions of $\tau_0$ and the power-law fitted curves for each parameter (black dashed lines, see the context), respectively.}
\label{simu}
\end{figure*}

Fig.~\ref{simu} shows results from our simulation.
Similar to $a=0.9$ cases, we observe the spark gap intermittently opens around the null surface.
This behavior is the result of the intermittent secondary pair production induced by the particle acceleration in the gap: particles accelerated in the gap up-scatter soft photons, which soon annihilate with other soft photons to produce electron-positron pairs.
Those secondary-produced pairs will compensate for the charge/current deficiency, and then after some time, the gap longitudinal electric field will be screened out.
The pair multiplicity in the gap affects the physical extension and the peak amplitude of the gap electric field, and as a result, $\gammae$, $\Lcur$ is sensitive to $\tau_0$.
$\Lic$ is less sensitive to $\tau_0$ than $\Lcur$ because (1) the persistent flow of plasma particles just outside of the gap is required to maintain the steady electric current, whose amplitude is independent to $\tau_0$, and then those particles will produce a steady flux of IC scattered photons, and (2) scatterings in the gap mainly occurs in the KN regime, in which IC emitting power is weakly dependent to the particle Lorentz factor $\gamma_{\rm e}$.
We obtain the averaged fitting function of those parameters: $\gammae\approx8.9\times10^6(\tau_0/30)^{-0.4}$, $\Lcur\approx10^{-2}(\tau_0/30)^{-2.8}\Lbz$, and $\Lic\approx5.8\times10^{-4}\Lbz\propto\tau_0^{-0.2}$ (black dashed lines in the figure).
$\Lcur/\Lbz$ and $\Lic/\Lbz$ have a weak dependence on the BH spin.
For simplicity, we assume that they are independent of $a$ in the context.
The factor 3-10 difference of $\Lcur/\Lbz,\,\Lic/\Lbz$ may result from the slightly larger size of the gap (top left panel of Fig.~\ref{simu}) leading to the larger amplitude of the gap electric field \citep[see Eq. (12) of][for a reference]{Kin24}, and also the weaker gravitational redshift effect for the lower spin value.

\bibliography{sample631}{}
\bibliographystyle{aasjournal}
\end{document}